%% file: 0_main.tex
\documentclass{article}

\usepackage{microtype}
\usepackage{graphicx}
\usepackage{subfigure}
\usepackage{booktabs} 

\usepackage{hyperref}

\usepackage[arxiv]{icml2024}

\usepackage{amsmath}
\usepackage{amssymb}
\usepackage{mathtools}
\usepackage{amsthm}
\usepackage{balance}

\usepackage[capitalize,noabbrev]{cleveref}

\theoremstyle{plain}

\theoremstyle{definition}

\theoremstyle{remark}

\usepackage[textsize=tiny]{todonotes}

\icmltitlerunning{Bringing Generative AI to Adaptive Learning in Education}

\begin{document}

\twocolumn[
\icmltitle{Bringing Generative AI to Adaptive Learning in Education}



\icmlsetsymbol{equal}{*}

\begin{icmlauthorlist}
\icmlauthor{Hang Li}{squiralai,msu}
\icmlauthor{Tianlong Xu}{squiralai}
\icmlauthor{Chaoli Zhang}{zjnu}
\icmlauthor{Eason Chen}{cmu}
\icmlauthor{Jing Liang}{squiralai}
\icmlauthor{Xing Fan}{squiralai}
\icmlauthor{Haoyang Li}{squiralai}
\\
\icmlauthor{Jiliang Tang}{msu}
\icmlauthor{Qingsong Wen}{squiralai}
\end{icmlauthorlist}

\icmlaffiliation{squiralai}{Squirrel Ai Learning by Yixue Education Inc.}
\icmlaffiliation{msu}{Michigan State University}
\icmlaffiliation{cmu}{Carnegie Mellon University}
\icmlaffiliation{zjnu}{Zhejiang Normal University}

\icmlcorrespondingauthor{Qingsong Wen}{qingsongedu@gmail.com}

\icmlkeywords{Machine Learning, ICML}

\vskip 0.3in
]



\printAffiliationsAndNotice{}  

\input{1_abstract}

\input{2_introduction}

\input{3_background}

\input{4_recent}

\input{5_bridge}

\input{6_discussion}

\input{7_conclusion}

\section*{Authors' Contributions}
Hang L. and QingSong W. conceptualized the manuscript. Hang L. and Tianlong X. created all figures and data visualizations. Hang L., Tianlong X., ChaoLi Z., Eason C. contributed to writing the manuscript, Qingsong W., Jiliang T., Jing L., Xing F., and Haoyang L. contribute to reviewing and editing the manuscript.

\section*{Competing interests}
The author declares no competing interests.






\balance
\bibliography{10_ref.bib}
\bibliographystyle{icml2024}


\input{appendix}


\end{document}

%% file: 1_abstract.tex
\begin{abstract}
The recent surge in generative AI technologies, such as large language models and diffusion models, has boosted the development of AI applications in various domains, including science, finance, and education. Concurrently, adaptive learning, a concept that has gained substantial interest in the educational sphere, has proven its efficacy in enhancing students' learning 
efficiency. In this position paper, we aim to shed light on the intersectional studies of these two methods, which combine generative AI with adaptive learning concepts. By presenting discussions about the benefits, challenges, and potentials in this field, we argue that this union will 
contribute significantly to the development of the next-stage learning format in education.
\end{abstract}

%% file: 2_introduction.tex
\section{Introduction}


Generative AI (GenAI) refers to artificial intelligence models that generate new content, including text, images, audio, and video, based on the patterns and information they have learned from the given training data \cite{cao2023comprehensive}. Unlike the traditional Machine Learning (ML) algorithms, which focus on analyzing and interpreting data, GenAI is designed to create new, original outputs and thus able to solve more challenging problems. With the recent success of GenAI technologies, such as large language models and diffusion models, in generating human-like outputs to a wide range of problems in Natural Language Processing (NLP) \cite{bubeck2023sparks} and Computer Vision (CV) \cite{ramesh2022hierarchical}, the development of applying GenAI to real-world problems is rapidly increasing, which spreads from science \cite{walters2020assessing, lopez2020enhancing}, finance \cite{rane2023role,bruhl2023generative} to education \cite{cooper2023examining, baidoo2023education}. 


Meanwhile, Adaptive Learning (AL), which is considered an emerging educational, technological innovation in education \cite{martin2020systematic}, has been demonstrated with pedagogical benefits, including acceleration, remediation, meta-cognition, mastery-based learning, immediate feedback, and interactive learning \cite{hattie2023visible}. Adaptive learning aims to generate unique learning experiences by accounting for individual differences to improve the scholastic path, learning process, and learner satisfaction in varied learning situations \cite{liu2017using}. As adaptive learning relies on collecting and analyzing data about learners' interactions, performance, preferences, and progress to tailor the educational experience to their individual needs, ML techniques have been extensively employed in the AL framework for supporting the analysis over large volumes of data about learners' behavior and assisting in creating personalized learning paths \cite{gheibi2021applying}. 

\textbf{Why This Position Paper?} 
The overwhelming performance of GenAI over ML algorithms in challenging data analysis problems provides new opportunities to current ML-based AL systems. By leveraging the advanced capabilities of GenAI in creating human-level responses to problems like reasoning, the new AL systems will be able to finish more complicated AL tasks, such as dynamic learning path planning. Although some pioneering works have started to explore the usage of GenAI for education purposes \cite{dan2023educhat,xiao2024automation}, there is no systematic review and discussion focusing on bringing GenAI to AL problems. As adaptive learning is increasingly recognized as a significant and promising direction for the future of education \cite{park2013adaptive}, in this position paper, we propose to review the related studies and call attention to the opportunities in combining GenAI with AL. Furthermore, by conducting comprehensive discussions on the benefits, challenges, and threats, we hope to shed light on future research in this direction and push AL into the next chapter.

\textbf{Our Contributions.} 
%
%
Our position in this paper is that bringing GenAI to AL will present both advantages and challenges to current ML-based AL systems, creating numerous opportunities and original topics for future educational development. To support this claim, we (1) conduct a comprehensive literature review on existing ML methods and GenAI works on adaptive learning, (2) explore the potential benefits and summarize ongoing industrial practices, (3) present in-depth discussions about the challenges and opportunities from the pedagogical perspective, and further extend these considerations to their broader educational impacts.

\textbf{Related Works.}
The impressive performance of GenAI, especially the Large Language Models (LLMs), in generating human-like responses to complicated requests \cite{bubeck2023sparks} encourages the recent research in exploring GenAI for education purpose \cite{dan2023educhat, kieser2023educational}. To provide researchers with a broad overview of the domain, numerous exploratory and survey papers have been proposed. For example, \citet{qadir2023engineering, rahman2023chatgpt} and \citet{rahman2023chatgpt} conclude the applications of  ChatGPT to engineering education by analyzing the responses of ChatGPT to the related pedagogical questions. \citet{jeon2023large} and \citet{mogavi2023exploring} collect the opinions from different ChatGPT user groups, e.g., educator, learner, researcher, through in-person interviews, online post replies and user logs, and conclude the practical applications of LLMs in education scenarios. \citet{baidoo2023education} and \citet{zhang2023systematic} focus on the literature review over the published papers and summarize the progress of the area with structured tables. Although the above works have covered a wide range of existing applications of GenAI in education scenario and provided their long-term visions for future studies, we argue that our position paper is still valuable as none of them focuses on the direction of bringing GenAI to AL. Compared to the general problems such as applying GenAI for education, we propose to present a more concentrated summarization and in-depth discussion toward the adaptive learning topic in this paper. 






%% file: 3_background.tex
\section{Background}

\subsection{Adaptive Learning (AL)}

\begin{figure}[!t]
    \centering
    \includegraphics[width=0.48\textwidth]{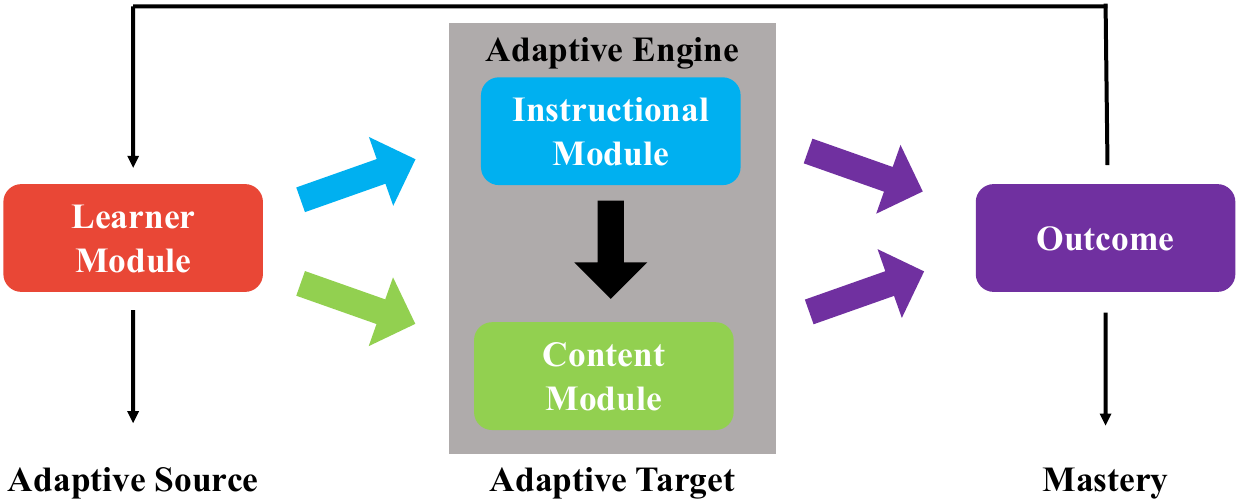}
    \hspace{-0.25cm}
    \caption{An Illustration of Adaptive Learning (AL).}
    \label{fig:al_framework}
\end{figure}

Teach-to-the-middle instruction is the most commonly used strategy for transitional teacher-led instructions as it benefits the majority of the students when the teacher resources are limited. However, such method does not fit learners who have different academic ability from the norm. To solve this problem, adaptive learning is developed, which aims to provide an efficient, effective and customised learning experience for students by dynamically adapting learning content to suit their individual abilities or preferences \cite{park2013adaptive, aleven2016instruction}. In the recent three decades, the effectiveness of AL has been widely recognized by the consistent superior performance reported in different researches' comparisons between AL users and the comparison sets \cite{mojarad2018studying, wang2023adaptive}. A valid AL system \cite{aleven2016instruction} usually consists of three components: learner module, content module, and instructor module, and its overview is shown in Fig.~\ref{fig:al_framework}.

\textbf{Learner Module}, also known as the student model, refers to the learner characteristics of what a student knows and does. The Learner Module includes learner attributes, learner preferences, learner knowledge and proficiency, motivational or emotional aspects of learner behavior, and individual differences that are used to adapt the learning.

\textbf{Content Module}, also known as the expert or domain model, refers to the content or knowledge base for the course. The Content Module could involve concepts that build on each other and include a learning map with relationships between different ideas and how the course content is delivered to the learner.

\textbf{Instructor Module}, also known as the pedagogical model, refers to the algorithm that assists in adapting the instruction based on the content and learner module, and defines what, when, and how adaptation can occur. Some of the adaptation techniques include pacing, the format of instruction, and sequencing. This model provides the base for deciding what content is presented to the learner and can also be called the adaptation model since it describes what is adapted and how it is adapted.

\subsection{Generative AI (GenAI)}

The origins of Generative AI can be traced back to the mid-1950s with the emergence of the concepts of artificial intelligence and machine learning. Unlike discriminative modeling, which has been the driving force of most progress in artificial intelligence, generative modeling problems are generally more difficult to tackle. Contributing to the rise of deep learning \cite{lecun2015deep}, Generative Adversarial Networks (GANs) \cite{goodfellow2014generative} and Variational Auto-encoders(VAEs) \cite{kingma2013auto} first revolutionized image generation and became a cornerstone in the field of GenAI. Moreover, the recent emergence of Transformer \cite{vaswani2017attention}, has marked another significant evolution in GenAI within the field of NLP, which has propelled GenAI into a new era. LLMs, such as OpenAI's GPT series, are the most representative ones of these GenAI models. By taking advantage of the vast number of parameters, extensive pre-training text corpus, and advanced fine-tuning methods \cite{ouyang2022training}, LLMs have shown impressive capabilities in understanding context, generating coherent and contextually relevant responses \cite{bubeck2023sparks}. Apart from that, further studies \cite{brown2020language} discover the strong few-shot (in-context) learning ability of LLMs, and such findings encourage the revolution in NLP solutions from task-specific methods to foundation models \cite{zhou2023comprehensive}. Despite the above progress, Generative Diffusion Models (GDMs) \cite{ho2020denoising, song2020score} also bring improvements to the generation quality of images \cite{dhariwal2021diffusion} and audio signals \cite{kong2020diffwave}. Finally, by connecting GenAI models from different modalities, multi-modal GenAI models are proposed and have achieved promising results in directions like text-to-image \cite{ramesh2022hierarchical}, text-to-audio \cite{popov2021grad} and text-to-video generations \cite{wang2023modelscope}.

%% file: 4_recent.tex
\section{Bringing GenAI to AL}

\begin{figure}[!t]
    \centering
    \includegraphics[width=0.45\textwidth]{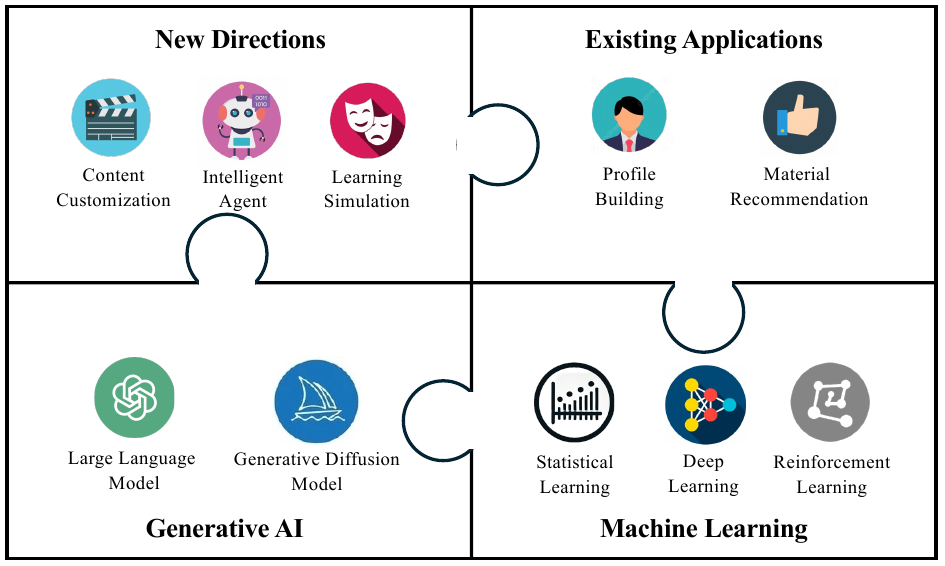}
    \caption{An overview of bringing GenAI to AL.}
    \label{fig:genaial_framework}
\end{figure}

In recent decades, ML plays a crucial role in AL by enabling the system to analyze learners' behavior, build learning profiles, and provide personalized adaptions to learners. In following sections, we summarize changes brought by GenAI from two perspectives: (1) Empower existing algorithms, (2) Establishing novel directions. The relationships between GenAI and adaptive learning are summarized in Fig.~\ref{fig:genaial_framework}.

\subsection{Empower Existing Algorithms}

\label{sec:existing}

Building students' learning profile and then recommend suitable study materials are core functions for AL systems. In this section, we present a brief summarizing about existing works, and explain from different perspectives that GenAI will be an complementary helper to existing algorithms.

\begin{figure*}[!t]
    \centering
    \includegraphics[width=0.9\textwidth]{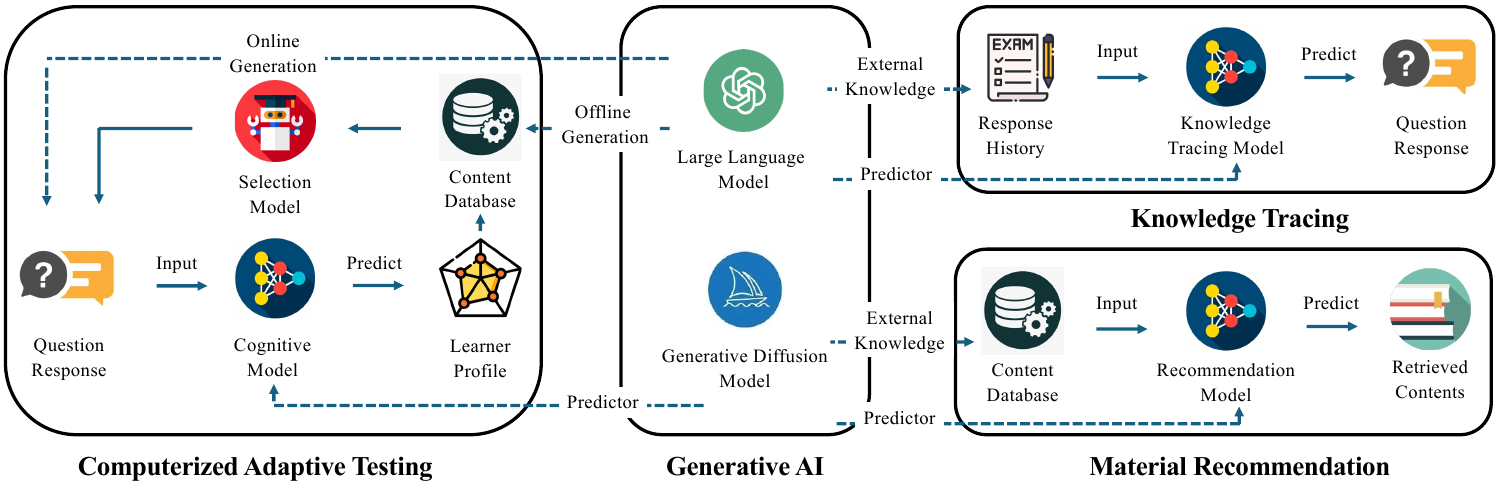}
    \caption{Empowering existing algorithms in AL system with GenAI.}
    \label{fig:existing_method}
\end{figure*}

\subsubsection{Profile Building}

The accurate profile information for different learner is an important foundation for any AL system. However, unlike profiles used in scenarios like commodity recommendation, profiles for adaptive learning focuses on tracking the learners' knowledge proficiency, which cannot be observed directly. To solve this problem, prior works developed Cognitive Diagnosis Models (CDM) on the basis of cognitive science or psychometrics \cite{ackerman2003using}, and exploit different types of ML method, such as statistical learning models \cite{hambleton1991fundamentals,de2009dina}, neural networks \cite{cheng2019dirt, wang2020neural}, to estimate the user's knowledge proficiency based on the interaction history between the examinees and question. Commonly, CDM is applied in traditional paper-and-pencil tests, however, the one-for-all approach dismiss the difference between individuals and usually faces to the challenges in efficiency and accuracy. To deal with that, Computerized Adaptive Testing (CAT) is propose, and it aims at tailoring the selection of questions to each examinee’s level of proficiency, thereby maximizing the accuracy of the assessment while minimizing the test length \cite{liu2024survey}. Due to the complexity of the optimization process, more advanced machine learning algorithms, such as reinforcement learning \cite{gilavert2022computerized}, active learning \cite{veldkamp2019robust,bi2020quality}, meta-learning \cite{ma2023novel,yu2023sacat}, are widely adopted by recent studies. Knowledge tracing (KT) is another popular research directions for learners' profiles building, which also uses machine learning methods to build learners' profiles. KT simplifies learner's proficiency estimation as his/her next question correctness prediction \cite{abdelrahman2023knowledge}, making it more flexible to diverse machine learning algorithms. The usual ML methods for KT include probabilistic graphical models \cite{corbett1994knowledge}, factor analysis models \cite{cen2006learning}, and deep learning sequential models \cite{piech2015deep}. 

The success of above algorithms demonstrates the effectiveness of ML algorithms in adaptive learning, on the basis of that, we position that introducing GenAI could further exploit the potential of existing ML methods. First, the human-instructor-level performance of GenAI in composing questions in subjects like math \cite{lee2024math}, computer science \cite{prokudin2023learning} and language studies \cite{xiao2023evaluating} are widely recognized by recent studies. Following explorations by \citet{adewumi2023procot} further verified the validness of using personalized question to enhance the engagement of student in test. These findings provide new opportunities for current CAT system in replacing pre-fixed question bank with dynamic generative pipelines, by which CAT can apply more flexible controls to questions and prior question bank exposure problem can be greatly resolved. Apart from that, leveraging GenAI to produce external knowledge for both KT and CAT data is also a valuable direction. As mentioned above, external knowledge like question knowledge concepts typologies is proven to be effective handlers in improving the performance of existing models. However, the external information requirement on data restricts the wide application of those new algorithms on different datasets. Contributing to powerful prior knowledge learnt by LLMs through the pre-training on large-scaled web corpus, using LLMs automatically add external knowledge, such as concept tags, to question becomes convenient, which will significantly encourage the emergency of stronger models on KT and CAT task. At last, by applying in-context learning (ICL) \cite{dong2022survey} and fine-tuning \cite{zhang2023instruction}, LLMs can also directly be used as predictors. Pioneering studies by \citet{neshaei2024towards} and \citet{jung2024clst} observe that LLM based predictor has great potential in solving the "cold-start" challenges for existing KT algorithms. Overall, the engagement of GenAI with existing profile building method brings new opportunities in the profile building researches of AL systems and the three above ways in leveraging LLMs for profile building is shown as Fig.~\ref{fig:existing_method}.

\subsubsection{Material Recommendation}

Learning material recommendation is another important component of AL systems, which provides different user with the customized learning materials. Based on optimization objects and considering factors, current approaches can be classified into different categories: content based, collaborative filtering based, knowledge based, and hybrid solutions \cite{khanal2020systematic}. Content-Based (CB) methods use machine learning models to generate informative attributes for each learner \cite{shu2018content}. Based on these attributes, a rule-based system will exclude the contents and recommend the materials from remaining results \cite{kolekar2019rule}. Collaborative Filtering (CF) make predictions about a learner's interests based on preferences from many learners. In addition to preference, learner style \cite{bourkoukou2017recommender} and skill levels \cite{han2016collaborative} are also considered by recent CF-based studies. Knowledge-based (KB) systems recommend items based on specific domain knowledge about how certain item features meet users’ needs and preferences \cite{aeiad2019adaptable, nitchot2019assistive}, and knowledge structures between learning contents are concerned by recent studies. At last, hybrid based solutions hybridizes the features of two or more above techniques to benefit from the strengths of each technique and to improve performance \cite{ghauth2010measuring}.

Although all above recommendation algorithms demonstrate good performance in their experiments, the application in real-world scenario still faces challenges, especially many of them require either delicate annotations or large scaled preference records \cite{khanal2020systematic}. In addition, as recent works prone to use deep learning models as their backbone model, many of these outputs are lack of intepretablity, which cannot provide valid support to learner's inquiry on recommended materials. Fortunately, the recent progresses in using GenAI as knowledge enhancer \cite{gong2023unified} and predictor \cite{cui2022m6} for recommendation system provides good demonstrations in solve the issues above. To be specific, GenAI models like LLMs' has powerful few-shot learning abilities, which enables it learn an effective predictor with limited annotated data \cite{brown2020language}. Besides that, LLMs' strong prior knowledge can be leveraged as external resource to enhance the inputs for all textual factors considered within the recommendation process \cite{li2023ctrl}. At last, the intermediate text before the outputs usually contain the reasoning process of LLMs, which automatically serve as good interpretation for the recommendation results \cite{chu2024llm}. Based on these facts, we position that the introduction of GenAI will empower existing material recommendation algorithms with stronger adaptation capabilities and the recommendation process will be more transparent and convincing to the users. The usages of GenAI in material recommendation are detailed in Fig.~\ref{fig:existing_method}. 

\subsection{Establishing  Novel Directions}

Apart from enhancing existing algorithms, GenAI's generative characteristic and its emerged planning and imitation capability establish novel directions for AL studies. In this section, we focus on three representative ones and introduce changes brought by GenAI based method in each direction.

\subsubsection{Content Creation}

The high fidelity of generated results, such as image \cite{betker2023improving}, video \cite{videoworldsimulators2024}, text \cite{achiam2023gpt}, demonstrates the great potential of GenAI in content creation. In addition, the emergent of techniques like instruction tunning \cite{zhang2023instruction} further facilitate the generating requirements by allowing users use verbal requirements to instruct LLMs during generation. \citet{cooper2023examining} explored the usages of GenAI models like GPT as instructor assistant to help create course outlines, compose a quiz focusing on some specific knowledge concepts, and even provide suggestion to teachers in organizing study plans for the specific lecture. \citet{hu2024teaching} leveraged GPT model to generate teaching plans in math for teachers and their evaluation results revealed the generated plans are achieving expert-level performance in multiple aspect, including problem chains organizations, teaching priority identifications and subject contents articulation. Besides textual learning contents, \citet{pierce20248} propose to use image-to-text generative model, DALLE-2 \cite{ramesh2022hierarchical}, to synthesis patient to educate doctors in training. By applying such approach, it will be possible to rapidly create photographic quality images without compromising patient confidentiality. 

\begin{figure}[!btph]
    \centering
    \includegraphics[width=0.38\textwidth]{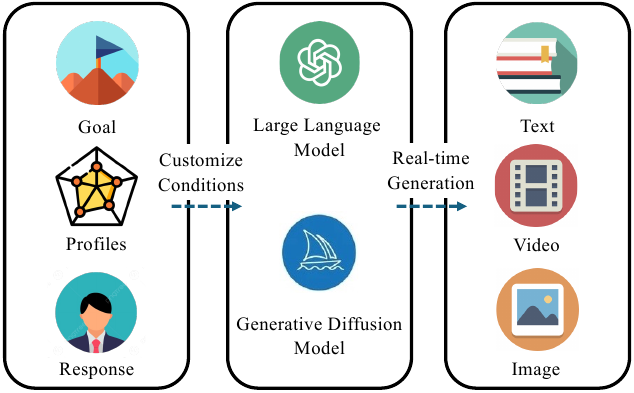}
    \caption{GenAI for Content Creation in AL.}
    \label{fig:content}
\end{figure}

As for adaptive learning, benefits brought by GenAI is not only limited to reduce the current workloads of instructors, but also provide new opportunities in creating learning contents aligning with learners' interest and proficiency dynamically \cite{pesovski2024generative}. Compared to static contents used by current AL systems, the dynamically created learning contents provide learner a more precised way in receiving knowledge. In addition, the real-time generation feature further empower AL systems to dynamically make adjustment based on changes happened in user's knowledge proficiency during the learning process. At last, the recent progress in mutlimodal GenAI research provides good foundations to incorporate various modalities into GenAI models \cite{bewersdorff2024taking}. In this case, the data formats of learning content will not be restricted by text only, the video and audio version contents can be helpful to improve the engagement of learner processes and provide convenient to people with special needs \cite{mallory2024empowering}. In a word, the emergence of GenAI is opening up new and creative ways to create contents for education and its usages for AL system are presented as Figure~\ref{fig:content}.

\begin{figure}[!btph]
    \centering
    \includegraphics[width=0.38\textwidth]{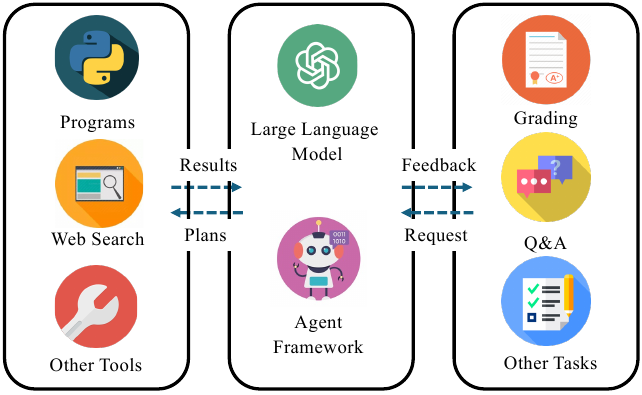}
    \caption{GenAI for Intelligent Agent in AL.}
    \label{fig:agent}
\end{figure}

\subsubsection{Intelligent Agent}

Using GenAI model, LLMs, as intelligent agent is a popular direction in recent AI communities \cite{wang2024survey}. The key idea of these researches is to leverage LLMs' task planning potentials to solve complicated real-world problems through its initiative elaborations with external resources such as computer programs, search engine and other toolkit. On the basis of that, many advanced LLM-agent frameworks are proposed and achieves promising results in various type of problems such as math problem solving \cite{wu2024mathchat}, program developing \cite{wu2023autogen} and question answering \cite{zhuang2024toolqa}. In adaptive learning scenario, the direct application of intelligent agent is personal learning assistant chat-bot \cite{zarris2023educational}, which focuses on provide real-time support to student during their study. Contributing to the strong instruction following capability of GenAI models like LLMs, student can post his question and request directly through verbal text, and GenAI will be able to leverage all available resources to generate the answers aligning the with student's needs. With incorporating techniques like prompt learning and fine-tuning, recent studies have successfully implemented learning chat-bots for essay writing \cite{han2024recipe4u} and program learning \cite{smith2024toward}. Apart from using it as tools only for student, agent-based framework can also be a powerful toolkit for instructors in adaptive learning. For example, providing personalized feedback on each student's assignment is a burdensome work for teachers. Prior studies tried to use machine learning algorithms to automate the procedure but only focus on specific question types \cite{fagbohun2024beyond}. Contributing to the exceptional tool using ability, GenAI based algorithms is able to provide valid solutions to question with more general formats. Besides that, the exception planning capability of LLMs discovered by recent studies \cite{huang2024understanding} also provides the vision that AL system can be organized and managed with the collaboration between different GenAI agents, and human instructors are free from the tedious routine works and focusing on providing supervisions on important decisions only.

\subsubsection{Learning Simulation}
 
Training models to produce indistinguishable samples with observed ones used to be a common training strategy for generative models like Generative Adversarial Networks (GAN) \cite{goodfellow2020generative}. As the most advanced generative model, GenAI models have been proven be able to produce high-quality imitations of specific styled images \cite{zhang2023inversion} and text corpus \cite{tao2024cat}. Moreover, contribute to the powerful in-context learning capabilities and conversational features of models like LLMs, using LLMs for role-playing and study the character of simulated roles has become an interesting topics in NLP communities \cite{shao2023character}. For adaptive learning, due to the privacy concerns and high collection costs, limited data size of student learning history is always the most common challenge faced by applying advanced machine learning models \cite{colchester2017survey}. With leveraging the role-play ability of GenAI, it becomes possible to generate large scaled synthetic dataset with low cost. Inspired by recent studies in computer vision domain \cite{yeo2024controlled}, we are promising to see the performance gain added to current method with incorporating synthetic student data during the training. Furthermore, contribute to the strong profile following characteristics of the LLMs-based role-play algorithm \cite{shao2023character}, using it to test outcomes for different learning plans on specific student will be possible, which helps both instructor and AL systems to find optimal solutions from a more objective perspective.

\begin{figure}[!btph]
    \centering
    \includegraphics[width=0.38\textwidth]{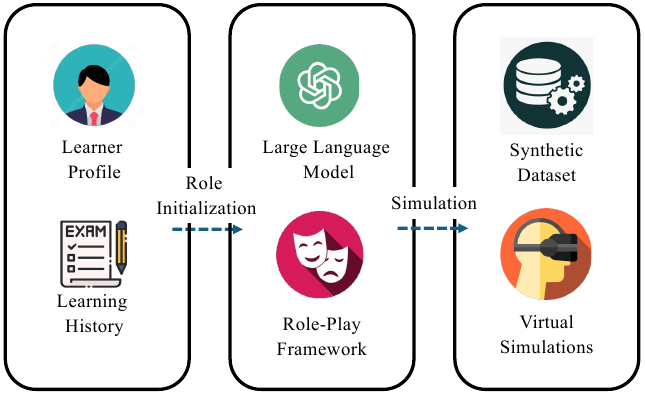}
    \caption{GenAI for Learning Simulation in AL.}
    \label{fig:agent}
\end{figure}

%% file: 5_bridge.tex
\section{Impacts of GenAI on AL}

In the following, we first outline advantages GenAI that can offer to AL, along with examples of industrial applications (Appendix.~\ref{app:industrial}). Then, we provide insights from an educational perspective, delving into the disadvantages and challenges faced by the new framework.

\subsection{Benefits and Strengths}

\subsubsection{Diversity and Dynamics}

Integrating GenAI into AL offers significant benefits, notably its ability to produce dynamic and diverse outputs. Current ML-based adaptive learning systems, despite their automation \cite{wang2023adaptive}, still fall short of replicating the nuanced interactions of a skilled human teacher. GenAI bridges this gap by generating real-time, tailored responses and learning plans based on each student's immediate questions and learning status. Moreover, the recent advancements in LLMs for context understanding \cite{brown2020language} further enhance the potential for providing personalized feedback, offering a substantial leap in optimizing learning outcomes. Apart from providing new features, GenAI holds the potential to enhance current ML models in AL systems. As detailed in Section~\ref{sec:existing}, profile building methods typically use learner's question practice records, but these are often pre-set and designed for mainstream students, potentially hindering accurate assessment of individual learning status. GenAI, with its dynamic and diverse capabilities, can generate questions tailored to each learner’s past performance, aiding in more effective profile building. 

\subsubsection{Multi-modality Convenience}

The powerful multi-modal processing capabilities enables GenAI to understand student's learning status in AL from a broader perspective. To be specific, by integrating GenAI, the AL systems will be able to interpret leaner emotions in speech and facial expressions together with the text input. The enhanced input information will help the following analysis of the student's learning characteristics such as learning style and learning status \cite{giannakos2023role}, which will also benefit the precise adaptions to the student's learning paths. Meanwhile, as GenAI's outputs can be multi-modal \cite{wu2023next}, the generated learning materials for knowledge concepts will become more inspiring. For example, with image generation models, the interpretations of geometry problems could be presented with figures or even a short animation, which will be easier for learners to understand compared to prior pure-text responses. 

\subsubsection{Powerful Prior Knowledge}

The scarcity of high-quality annotated data in education is the challenge faced by many ML methods in AL system. And tt is caused by the strong privacy properties and high annotation costs of pedagogical data. The appearance of GenAI models brings new solutions as GenAI models are proven to be general-purpose foundation models \cite{bruhl2023generative}. Contributing to the powerful prior knowledge and \cite{dang2022prompt} and in-context learning capability \cite{dong2022survey}, existing algorithms in AL system are expected to receive a performance boost after the integration it with the prior knowledge provided by GenAI. Besides that, some new tasks with very few annotated data will also be able to get launched through GenAI's zero-shot or few-shot learning algorithms \cite{brown2020language}. This change will encourages the further explorations of using ML to solve more challenging tasks in AL. At last, the emergent planning prior knowledge enables GenAI not only be a specific executor but a general task planner. It open a door for developing intelligent agent in solving problem through the elaboration of external tools and resources. On the basis of that, some previous complicated tasks in AL will get solved soon. 

\subsection{Disadvantages, Challenges and Potentials}

\subsubsection{Hallucination}
The biggest issue with GenAI currently is its potential to produce content that doesn't actually exist, a phenomenon known as "Hallucination" \cite{ji2023survey}. Hallucinations happened even in the most advanced GenAI models. Research suggests that educating students about the limitations of using GenAI is crucial, emphasizing its susceptibility to errors when working with it \cite{strzelecki2024investigation}. This brings forth numerous opportunities and research questions for the future, such as how to design the AL system to help students cultivate an appropriate reliance on AI-generated content \cite{vereschak2021evaluate}.  Furthermore, hallucination could become a new adaptive teaching method, such as creating new and possibly erroneous educational content and asking students to practice the role of a tutor in correcting AI's mistakes \cite{ma2023hypocompass}.

\subsubsection{Capability Decay}
The ease of obtaining information from GenAI can create a dependency, where students might prefer quick GenAI answers over the process of learning and discovery. This over-reliance could erode critical thinking skills and reduce the students' capacity for self-driven inquiry. One potential solution is to integrate GenAI as a tool that prompts further questioning and exploration rather than providing definitive answers. Encouraging projects and assignments that require independent research and critical analysis can also help maintain a balance between utilizing GenAI and fostering self-reliance. However, these suggestions only begin to address the broader challenge of integrating GenAI into education without compromising the development of independent learning skills. This issue remains an open question, presenting a wide array of opportunities for educators, technologists, and policymakers to explore innovative solutions that balance the benefits of GenAI with the imperative of nurturing autonomous, curious learners.

\subsubsection{Fairness}

Fairness is a crucial ethical concern, particularly in education where GenAI technologies are being integrated \cite{fenu2022experts}. These advancements, while beneficial, bring to light issues of equitable access and unbiased treatment among learners. For instance, discrepancies between those accessing advanced systems like GPT-4 and those with older versions like GPT-3.5 can lead to unequal knowledge distribution, potentially exacerbating educational gaps. The effective deployment of GenAI in education also hinges on user familiarity and accessibility. Students well-acquainted with AI technologies stand to gain more, while those less experienced may encounter challenges and a steeper learning curve. Introducing standardized training programs might help bridge this gap, though such measures require meticulous implementation and oversight. Additionally, inherent biases within GenAI pose a significant challenge. These systems, limited by the biases in their training data, might offer skewed guidance, making vigilant monitoring essential to ensure fairness. In adaptive learning, where GenAI tailors instructional guidance, the risk of bias amplifies. Personalizing learning experiences is beneficial but needs to be weighed against the risk of unfairness. Ensuring an optimal, equitable, and unbiased educational experience for every learner involves navigating these complexities, making the balance between personalized learning and fairness both a technical hurdle and an ethical necessity in the integration of GenAI into educational systems.

\begin{figure}
    \centering
    \includegraphics[width=0.47\textwidth]{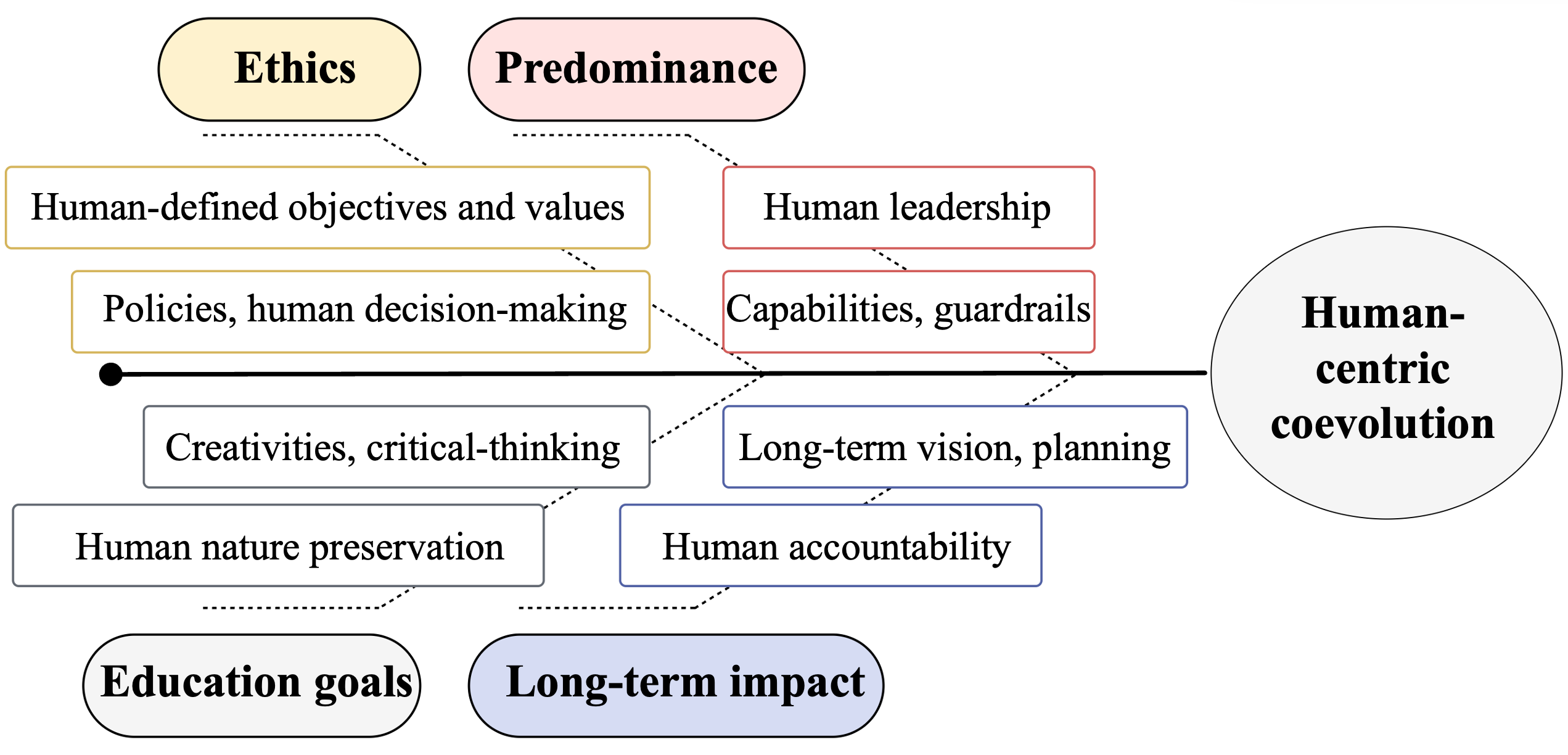}
    \caption{Human-centric Coevolution of GenAI and Education}
    \label{fig:opportunities diagram}
\end{figure}

\subsubsection{Coevolution of human, GenAI, and education} 

The realm of GenAI evolves as an indispensable tool in the educational sphere. This progression, however, underscores a nuanced balance: GenAI's development must be aligned with, but not overshadow, the broader trajectory of human advancement. Generative AI should serve as an augmentative force, enhancing human capabilities rather than supplanting them. The essence of this coevolution lies in the acknowledgment that GenAI, while a profound enabler, should not assume a dominant or leadership role. Generative AI's integration into educational systems should amplify human potential without dictating the course of our development. 
As shown in Fig.~\ref{fig:opportunities diagram}, we  consider the above issues from the following perspectives.

\textbf{Governance and Ethical Frameworks:} How do we establish robust governance structures and ethical frameworks that ensure generative AI's role in education adheres to human-defined objectives and values? This involves creating policies that prioritize human decision-making in critical educational processes, thereby preventing generative AI from autonomously setting or altering educational agendas.

\textbf{Human predominance:} In what ways can we design generative AI systems that inherently support and enhance human leadership in educational settings? How do we strike a balance between leveraging generative AI's capabilities for autonomous functioning and maintaining human control over its applications? This balance is crucial to ensure that while generative AI can operate efficiently in certain aspects of education, it does so under the guidelines and constraints set forth by human educators and policymakers.

\textbf{Educational Goals and generative AI Limitations:} In defining educational goals, how to ensure the core of education remains intrinsically human – an arena where human intellect, values, and creativity guide our journey towards a more enlightened future? It's essential to outline the boundaries of generative AI in handling complex, nuanced aspects of education that are inherently human.

\textbf{Long-term Impact and Accountability:} What mechanisms should we implement to regularly assess the long-term impact of generative AI in education, ensuring that it aligns with human-centric goals and adapting strategies as needed? This involves establishing accountability systems where a golden rule lies on the fact that humans always be the only accountable party that is responsible for the critical decisions, the road map outlined, as well as all outcomes.

%% file: 6_discussion.tex
\section{Further Discussion}

It's understandable that some readers might have different opinions toward our claims in this paper for a range of reasons. Below, we explore a few of these perspectives and share our discussions in an impartial manner:

\textbf{Why is our emphasis on adaptive learning rather than other educational perspectives?} 

Our response to this inquiry can be succinctly encapsulated from two angles: (1) adaptive learning in the educational sphere has a well-established history spanning several decades. Its efficacy in aiding learners to attain enhanced educational outcomes has been consistently validated through numerous studies. By integrating GenAI with adaptive learning, our goal is to maintain and accelerate the evolution of adaptive learning, propelling it into a new era of advancement; (2) the inherently data-driven nature of adaptive learning fosters the widespread adoption of machine learning algorithms within this field. Drawing on insights and methodologies from other ML-centric areas, such as NLP and CV, we are confident that GenAI will serve as a potent catalyst, steering adaptive learning in novel and exciting directions.

\textbf{What is the scope of this work?}

In this paper, we primarily focus on our reviews and discussions of ML-related components in AL systems, particularly because of their close relationship with GenAI. However, just as the example of "unconventional" solutions advised by GenAI, the deeper explorations in applying GenAI could bring us new visions to educational paradigms. Although this paper might not encompass all these emerging opportunities and challenges, its value remains significant. It aims to illuminate a promising direction for future research and development in this field. 

\textbf{How does adaptive learning combined with generative AI broadly impact the field of education?}

The integration of GenAI with adaptive learning enhances a previously practical tool, making it more efficient. This advancement not only improves educational methods but also illuminates potential pathways for further revolutions in education technology, particularly concerning universal challenges and risks associated with GenAI models. Moreover, the robust applicability of this method raises the possibility of addressing disparities in educational resources, potentially offering more substantial assistance in this area.

%% file: 7_conclusion.tex
\section{Conclusion}

In this paper, we aim to draw attention on the novel research area of integrating generative AI with adaptive learning in education. We believe that the data-centric nature of adaptive learning, which already embraces machine learning, provides an ideal platform for generative AI to enhance the efficacy of existing AL algorithms through its dynamic and diverse output capabilities. However, we also acknowledge the challenges and uncertainties that GenAI introduces to adaptive learning, particularly in sensitive areas like fairness and reliability. Drawing inspiration from GenAI's advancements in other fields, we remain confident that these concerns will be addressed, transforming these challenges into research opportunities and objectives. While our perspectives might not resonate with every reader, our aim is to spark conversation and encourage exploration in this field. If this paper can inspire future research, then it has successfully served its purpose.

%% file: appendix.tex
\newpage
\appendix
\onecolumn
\section{Appendix.}
\label{app:industrial}



In table below, we summarize those pioneering institutions which has begun their industrial practice with bringing generative AI (GenAI) to adaptive learning. The majority of these business cases focus on leveraging GenAI to empower their original products and creating enhanced personalized learning experience to their customers.

\begin{table*}[!btph]
  \centering
  \caption{Industrial practice case studies. USP is the unique selling proposition of the product.}
  \label{tab:indust_pract}
  \begin{tabular}{|p{0.1\textwidth}|p{0.15\textwidth}|p{0.68\textwidth}|}
  \hline
    Company & USPs & Product features\\
    \hline
    \hline
    \hyperlink{https://www.duolingo.com/}{Duolingo 
    } & Personalized roleplay & 
    Uses LLMs to deliver adaptive, highly personalized interactions, closely mimicking real-life conversations. Offers detailed AI-powered post-conversation feedback focusing on the accuracy and appropriateness of responses, identifying strengths and areas for improvement. 
    \\
    \hline
    \hyperlink{https://www.khanacademy.org/khan-labs}{Khanmigo 
    } & 1 on 1 Tutoring & Offers a virtual coach, adept at guiding students through problem-solving processes with carefully crafted probing prompts. 
    This method ensures that the learning process 
    is interactive and responsive. 
    \\
    \hline
    \hyperlink{https://www.adaptemy.com/}{Adaptemy 
    } & Learning content recommendation & Utilizes a triad of models—curriculum, content, and learner to deliver personalized learning experiences, enabling real-time adaptation of content, assessments, and learning paths based on performance.
    \\
    \hline
    \hyperlink{https://squirrelai.com/}{Squirrel Ai 
    }& Personalized learning planning & Established a large adaptive models (LAM) framework 
    to link the learning contents knowledge graph, student profiles and a study-path planning recommendation system to provide personalized learning paths planning. 
    \\
    \hline
    \hyperlink{https://www.dreambox.com/admin/solutions/adaptive-learning}{DreamBox Learning}& Math adaptive learning & Offers personalized K-8 math learning with supplemental curriculum and a range of live and self-paced resources to meet the learning needs of entire staff.\\
    \hline
    \hyperlink{https://www.chegg.com/}{Chegg} & 24/7 study support & Provides reliable solutions backed by experts, step-by-step explanations and learn tailorings. \\
    \hline
    \hyperlink{https://www.carnegielearning.com/}{Carnegie Learning} & K–12 education & Offers clear solutions family and has been proven to deliver up to 2x performance improvement on standardized tests. \\
    \hline
    \hyperlink{https://kahoot.com/}{Kahoot!}& Learning games & Provides learning games to make learning fun, easy, and rewarding for everyone.  \\
    \hline
    \hyperlink{https://en.100tal.com/}{TAL Education Group} & Smart learning solutions & Offers comprehensive learning services to students from all ages through diversified class formats. 
    \\
    \hline 
  \end{tabular}
  \label{industrial-practice}
\end{table*}